\documentstyle[aaspp4,11pt,epsf]{article}
\def\sun{\odot}
\def\lsim{\, \lower2truept\hbox{${< \atop\hbox{\raise4truept\hbox{$\sim$}}}$}\,}
\def\gsim{\, \lower2truept\hbox{${> \atop\hbox{\raise4truept\hbox{$\sim$}}}$}\,}

\begin{document}

\title{Local Luminosity Function at 15$\mu m$ and Galaxy Evolution Seen
by ISOCAM 15$\mu m$ Surveys} 
\author{Cong Xu}
\affil{Infrared Processing and Analysis Center, Jet Propulsion Laboratory, 
Caltech 100-22, Pasadena, CA 91125}
\received{Nov. 30, 1999}
\accepted{ ... }

\begin{abstract}

A local luminosity function at 15$\mu m$ is derived using
the bivariate (15$\mu m$ vs. 60$\mu m$ luminosity) method,
based on the newly
published ISOCAM LW3-band (15$\mu m$) survey of the very
deep IRAS 60$\mu m$ sample in the north ecliptic pole
region (NEPR). New IRAS 60$\mu m$ fluxes are obtained using 
the SCANPI/SUPERSCANPI software at the new ISOCAM positions of 
the sources in the sample. It is found to be in excellent agreement
with the 15$\mu m$ local luminosity function published by
Xu et al (1998), which is derived from the {\it predicted}
15$\mu m$ luminosities of a sample of IRAS 25$\mu m$ selected
galaxies. Model predictions of number counts and redshift 
distributions based on the local luminosity
function and assumptions of its evolution with the redshift
are calculated and compared with the data of ISOCAM 15$\mu m$ surveys.
Strong luminosity evolution on the order of $L\propto (1+z)^{4.5}$
is suggested in these comparisons, while pure density evolution can
be ruled out with high confidence. The sharp peak at about 0.4mJy
in the Euclidean
normalized differential counts at 15$\mu m$ can be explained
by the effects of MIR broadband emission features, eliminating
the need for any hypothesis for a 'new population'. 
It is found that the contribution
from the population represented by
ISOCAM 15$\mu m$ sources can account for
the entire IR/submm background, leaving little room
for any missing 'new population' which can be
significant energy sources of the IR/submm sky.

\end{abstract}


\section{Introduction}

Our understanding of galaxy formation and evolution in the early
epochs of the Universe has been vastly improved in the past a few
years, thanks mainly to new deep surveys in a wide range of wavebands,
ranging from the HST's WFPC2 (UV and optical) and NICMOS (NIR) surveys
in the Northern and Southern Hubble Deep Fields (Williams et al. 1996;
Williams et al. 1998; Thompson et al. 1999) to the SCUBA submm surveys
(see Blain et al. 1999b for a review). In particular, several
mid-infrared (MIR) and far-infrared (FIR) deep surveys (see Elbaz et
al. 1998b, Puget \& Lagache 1998 for reviews) were conducted using the
Infrared Space Observatory (ISO) (Kessler et al. 1996) during its 29
months mission (Nov. 1995 -- April. 1998).  The new results from these 
surveys (Aussel et al. 1999; Puget et al. 1999; Dole et al. 1999)
indicate significant improvements in sensitivity and accuracy over the
earlier published results (Rowan-Robinson et al. 1997; Kawara et
al. 1998). Strong cosmic evolution in the population of
infrared-emitting galaxies is indicated in these results
(Rowan-Robinson et al. 1997; Kawara et al. 1998; Aussel et al. 1999;
Puget et al. 1999), consistent with the results of SCUBA surveys
(Blain et al. 1999a) and with the scenario hinted at by the newly
discovered cosmic IR background (CIB) (Puget et al. 1996; Hauser et al. 
1998; Dwek et al. 1998; Fixsen et al. 1998).
These results challenge those from the
UV/optical surveys (Madau et al. 1998; Pozzetti et al. 1998) in the sense
that substantially more (i.e. a factor of 3 -- 5) star formation 
in the earlier Universe with $z\gsim 2$ may be hinted at in the IR/submm 
counts and in the CIB (see, e.g. Rowan-Robinson et al. 1997)
compared to that derived from 
the UV/optical surveys (Madau et al. 1998; Pozzetti et al. 1998).
The reason of this discrepancy is attributed to dust extinction
which may hide much of the star formation in the early Universe
from the UV/optical surveys (see Lonsdale 2000 for a review).

The best observed band in these ISO surveys is the ISOCAM LW3
($15\mu$m) band, with 14 surveys covering a wide range of flux density
from 0.05 mJy to 50 mJy (Elbaz et al. 1998b).  Compared to the longer
wavelength ISOPHOT surveys (e.g. 175$\mu m$ FIRBACK survey, Puget et
al. 1999), ISOCAM LW3 surveys have the advantage of using large
detector arrays ($32\times 32$ compared to the $2\times 2$ array of
the ISOPHOT-C200 camera) and having much better angular resolution
which allowed the surveys to go very deep ($\sim 0.1$ mJy compared to
sensitivity limits of $175\mu m$ surveys of $\sim 100$ mJy) before
reaching the confusion limit.  In addition to the indication of the
significant cosmic evolution of IR galaxies, two other very
interesting results are emerging from these MIR surveys: 
\begin{enumerate}
\item The
Euclidean normalized differential number counts of the 15$\mu m$ band
have a sharp peak at about 0.4 mJy (Elbaz et al.  1999)
which, Elbaz et al. (1998a) claim, can only be explained by 
adding to the number count model a new population of objects which
emerge (with increasing z) 
rapidly after $z\sim 0.4$ and start to dominate the counts below 1 mJy, 
but contribute negligibly in brighter flux range (Fig.10 of Elbaz
et al. 1998a). 
\item Subject to the substantial uncertainties in their IR SEDs, the
integrated light of the galaxies detected in 15$\mu m$ surveys may
account for most the IR/submm background (Elbaz et al. 1999).
\end{enumerate} 
Taken at face values these results may have far-reaching impact
on the studies of galaxy evolution, suggesting that the 
objects most responsible for the CIB, which are mostly missed 
by the UV/optical surveys, are already identified in the 
15$\mu m$ surveys and, perhaps 
more interestingly, they are from a new population
not seen in the local Universe (i.e. not among the IRAS sources which
in general have $z\lsim 0.1$). 

However, scrutiny of these interpretations of the
15$\mu m$ counts is imperative
because of the following two complications: (1) the effects of prominent 
emission features in the wavelength range of 3 -- 13$\mu m$
(Puget and Leger 1989) which can cause very significant K-corrections
in MIR surveys (Xu et al. 1998); and (2) the lack of a local 
luminosity function (LLF) in the 15$\mu m$ band which is needed for
the quantitative determination of the evolution rate from the number counts
(to date, the IRAS 12$\mu m$ luminosity functions have been mostly used 
in the interpretation of the ISOCAM 15$\mu m$ counts, resulting in
large uncertainties due to the significant 
variations of the $f_{15\mu m}/f_{12\mu m}$ among galaxies; see
Elbaz et al. 1999). Indeed, the source count model of Xu et al. (1998),
which takes into account the effect of the MIR emission features,
did predict the bumps and dips in the counts similar to what is seen in the
15$\mu m$ surveys. Accordingly, Xu et al. (1998) made the warning
that determinations of the evolution rate based on the slope of source counts
will have to treat the effects of the MIR emission features 
carefully. Xu et al. (1998) also derived a LLF at 15$\mu m$
from the {\it predicted} 15$\mu m$ flux densities of a 
sample of 1406 IRAS 25$\mu m$ selected galaxies based on a three
component (cirrus/PDR, starburst, AGN) SED model.

Recently, Aussel et al. (2000) published an ISOCAM 15$\mu m$ survey of
the very deep IRAS 60$\mu m$ sample in the northern ecliptic pole region 
(NEPR) (Hacking and Houck 1987, hereafter HH87; Hacking 1987). 
In this paper, we will derive a new 15$\mu m$ LLF
based on the data of this survey, and on the new IRAS SCANPI
measurements at 60$\mu m$ of the same sources using the new ISOCAM positions.
This is then compared to the LLF of Xu et al. (1998). With the confidence
in the 15$\mu m$ LLF gained from this new study we explore further,
using the model of Xu et al. (1998), the quantitative interpretation of
results of the ISOCAM deep surveys as published by Elbaz et al. (1999) and by
Aussel et al. (1999).

\section{New 15$\mu m$ LLF from NEPR Sample}
\subsection{New IRAS SCANPI measurements at ISOCAM positions}
The ISOCAM observations at 15$\mu m$ for 94 out of 98 galaxies in the 
very deep IRAS 60$\mu m$ sample in the NEPR (HH87)
are described out by Aussel et al. (2000). Altogether 106 sources
were detected with signal to noise ratios $\geq 3$. 
Several IRAS sources correspond to multiple ISOCAM
sources, given the much better angular resolution of ISOCAM ($\sim 10''$)
compared to the IRAS resolution ($> 1'$). The mean position offset
of the ISOCAM sources relative to IRAS positions (HH87)
is $\sim 10''$. 
\begin{figure}
\plotone{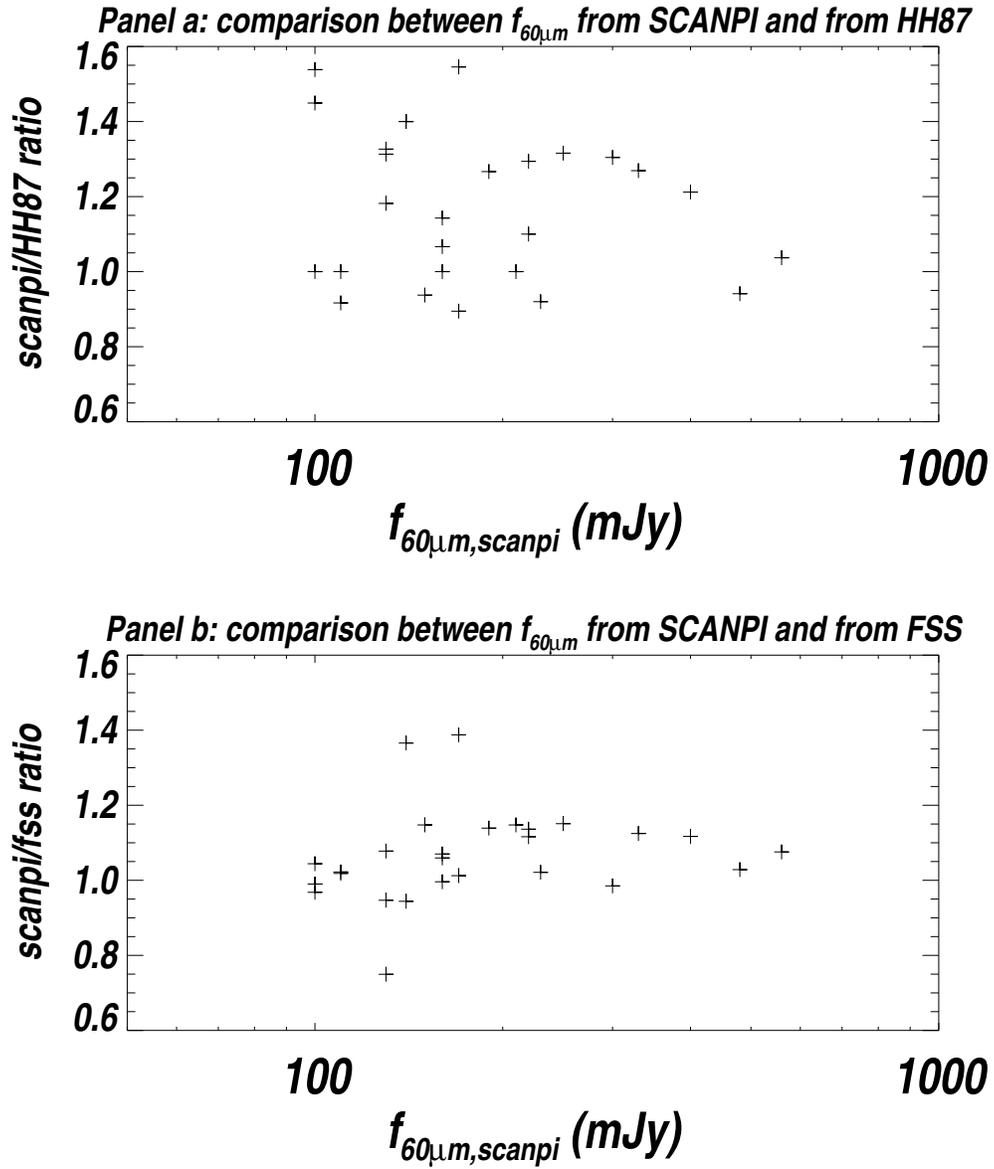}
\vskip-2truecm
\caption{
Comparisons between $f_{60\mu m}$ from
this work (new SCANPI/SUPERSCANPI reductions)
and from HH87 (Fig.1a), and from FSS (Fig.1b).}
\end{figure}

New IRAS measurements at 60$\mu m$ using the IPAC software SCANPI
were carried out at the positions
of the 117 (11 having signal to noise ratio $<3$)
ISOCAM sources in the NEPR sample listed in the Table 3 of
Aussel et al. (2000), exploiting the IRAS survey database.
This is to get a better
corresponding $60\mu m$ flux density for each ISOCAM source, and also to bring
the IRAS flux densities obtained by HH87 to the newest
IRAS standard (Moshir et al. 1992). 
For this task, three SCANPI queries were run and finished
on Oct. 28, 1999, using the default settings of 
SCANPI\footnote{For details of SCANPI and SUPERSCANPI processing,
see the webpage \newline
\centerline{
$http://www.ipac.caltech.edu/ipac/iras/scanpi$\_$over.html$.      }}.
Of the 117 sources, 99 are detected at 60$\mu m$.
For the 18 ISOCAM sources undetected by SCANPI, the interactive SUPERSCANPI
has been run in attempts to increase the sensitivity when,
in addition to the survey data, data from pointed observations
(HH87; Gregorich et al. 1995)
are also available. As in HH87 we include 
only pointed observations with 'deep-sky' macros in the SUPERSCANPI 
processing. Five additional detections were obtained through the
SUPERSCANPI coadds. This leaves 13 ISOCAM sources undetected in 
the 60$\mu m$ band, which have in general large offsets 
($<$offset$> \sim 60''$) from the original IRAS positions.
Of the 104 sources detected by SCANPI and SUPERSCANPI in the
60$\mu m$ band,
many (usually those corresponding to the same source in the HH87 
list) are confused. The plots of SCANPI and SUPERSCANPI
processing of these confused sources awere manually inspected to
determine the best total flux densities of these confused sources, which are
assigned to them jointly. A final list of 106 sources is given in Table 1,
including the sources undetected by SCANPI/SUPERSCANPI, 
and the 4 sources for which Aussel et al. reported only upper-limits
at $15\mu m$. Redshifts are found for 68 of them from Ashby et al. (1996).

The SCANPI/SUPERSCANPI results are compared with the flux densities
reported by HH87 and with those from the IRAS Faint Source
Survey (FSS) (Moshir et al. 1992) for sources which observe the
following criteria: (1) ISOCAM position within
20$''$ of the IRAS position, (2) not confused with any other
sources in the IRAS 60$\mu m$ band, and (3) for FSS comparison,
they have to be listed in the 
Faint Source Catelog (FSC) or in the Faint Source Reject File (FSR)
in the IRAS database. Thirty sources are selected by
the first two criteria, 27 of them also pass the criterion (3).
It is found (Fig.1a and 1b) that
the 60$\mu m$ flux densities obtained in this paper are consistent with
those listed in the FSC and FSR, but about 20\% higher than those 
reported by HH87. Given that the FSC and FSR results represent the
new standard of IRAS products, it is likely that the lower
flux densities of HH87 are due to some systematic biases
in the early processing of IRAS data.

\subsection{New 15$\mu m$ LLF Derived from Bivariate 15$\mu m/$60$\mu m$ LF}
Exploiting the NEPR sample, a LLF at 15$\mu m$ can be constructed 
using the so called 'bivariate method' (see, e.g., Xu et al. 1998), 
transferring the 60$\mu m$ LLF
of IRAS galaxies, which has been well studied in the literature 
(Soifer et al. 1987; Saunders et al. 1991; Yahil et al. 1992), to 
15$\mu m$ LLF utilizing the $L_{15\mu m}/L_{60\mu m}$ ratio
v.s. $L_{60\mu m}$ relation. We include only the 64 sources in Table~1 
which are detected in the 60$\mu m$ band (including sources
with upper-limits at 15$\mu m$) and which have measured redshifts
(Ashby et al. 1996). 
\begin{figure}
\plotone{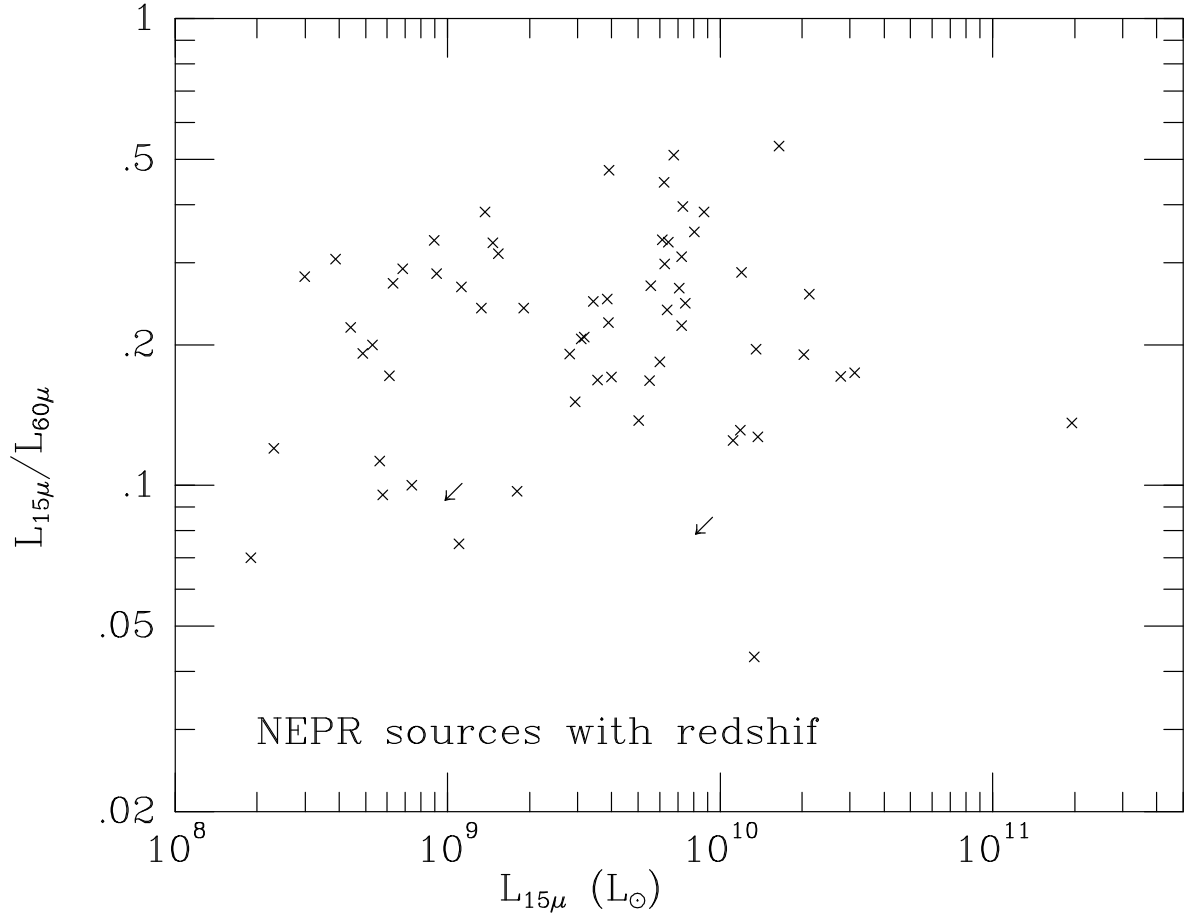}
\vskip-8truecm
\caption{
NEPR sources (64) used in the bivariate analysis:
$L_{15\mu m}/L_{60\mu m}$ ratio versus 15$\mu m$ luminosity.
}
\end{figure}

There are several concerns with regard to the sample:
\begin{itemize}
\item The sample is incomplete. In particular, the requirement of 
redshift availability excludes about one third of the sources
in Table 1.
\item Possible misidentification between the sources in the redshift survey
of Ashby et al. (1996) and the sources in this work.
\item Given the depth, the sample is not really local (many sources have
$z > 0.1$), hence may be affected by galaxy evolution with increasing redshift.
\item The redshift distribution of the sources shows strong clustering
(Ashby et al. 1996).
\end{itemize}
Will these affect the bivariate 15$\mu m/60 \mu m$ luminosity function?
The answer to this question depends on whether the 
15$\mu m$-to-$60 \mu m$ color ratio is a sensitive function of the luminosity.
This is because all the above potential problems with the sample are related 
to the redshift, and result in uncertainties in the luminosity distribution
(the 'visibility function') but not in the $L_{15\mu m}/L_{60\mu m}$ ratio
distribution. If the color ratio is insensitive to the luminosity,
the conditional probability function 
$\Theta(L_{15\mu m}/L_{60\mu m}|L_{60\mu m})$ (cf. Eq(7) of Xu et al. 1998),
which converts the 60$\mu m$ LLF to the 15$\mu m$ LLF, will be rather constant 
and won't be affected significantly by uncertainties associated with 
the luminosity.

Indeed, as plotted in Fig.2, the $L_{15\mu m}/L_{60\mu m}$ ratio of
sources in the NEPR sample appears to be rather
insensitive to the luminosity. This result is similar to
that of Soifer and Neugebauer (1991) who found that the 
$L_{25\mu m}/L_{60\mu m}$ ratio does not depend on the infrared luminosity.
This may not be surprising given that
the mechanisms of the 15$\mu m$ emission and of the 25$\mu m$ emission 
are nearly the same, namely due dominantly to the emission of
small grains undergoing temperature fluctuations in normal galaxies such as 
the Milky Way, and to the warm dust emission associated with 
star formation regions in starburst galaxies such as M82 
(D\'esert et al. 1990). This dualism in the radiation mechanism of
the MIR continuum, at the wavelengths not contaminated by the MIR emission
features, is the major reason for the lack of dependence of the two 
color ratios ($L_{15\mu m}/L_{60\mu m}$ and  $L_{25\mu m}/L_{60\mu m}$)
on luminosity. At the same time, the very cold  $L_{25\mu m}/L_{60\mu m}$
ratios of ultraluminouse galaxies (ULIRGs), which are due mostly 
to extinction at $25\mu m$ (Xu and De Zotti 1989), also further weaken
any statistical dependence of the $L_{25\mu m}/L_{60\mu m}$ ratio on the
grain temperature, the latter is a strong function of the luminosity
as demonstrated by the $L_{60\mu m}/L_{100\mu m}$ v.s. luminosity relation
(Soifer and  Neugebauer 1991).
\begin{figure}
\plotone{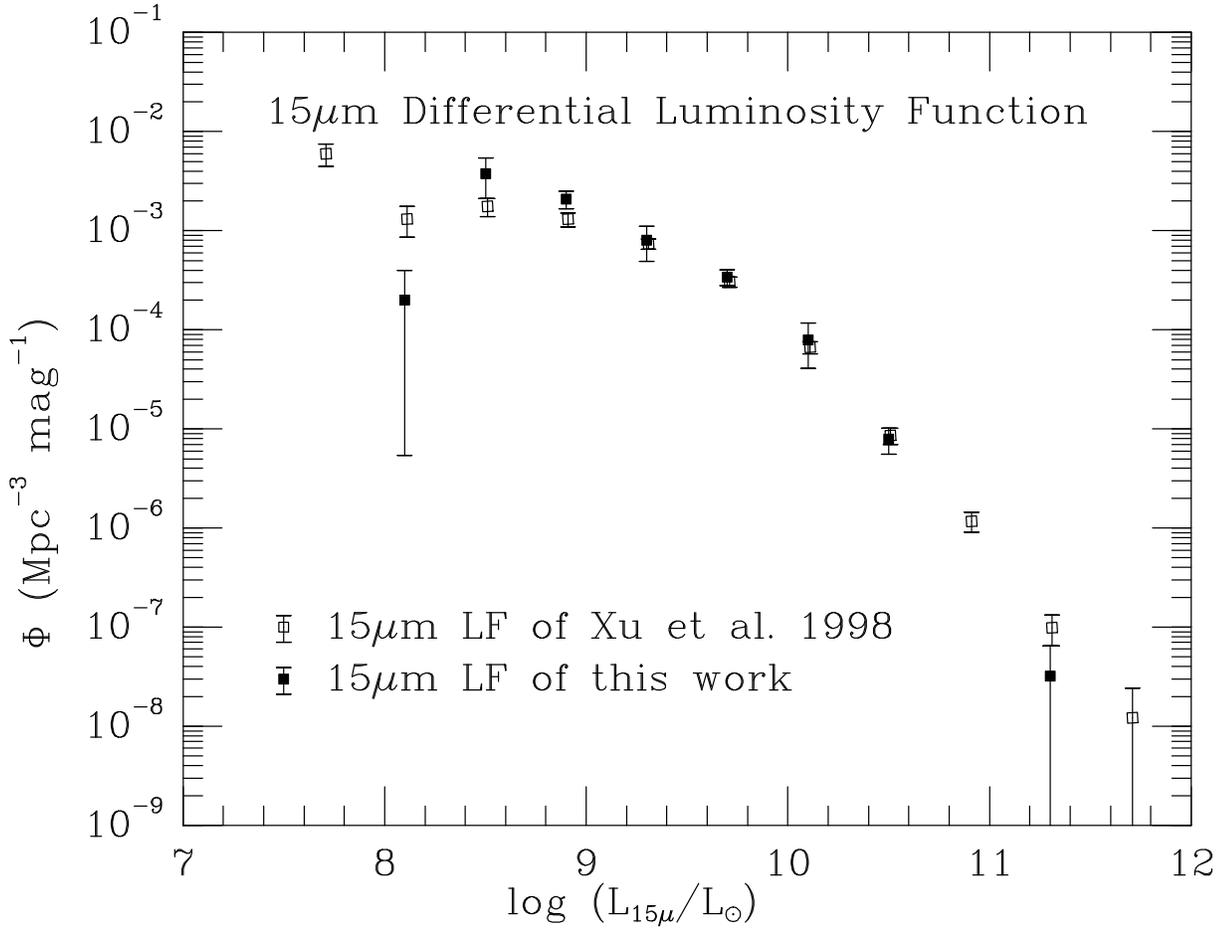}
\vskip-8truecm
\caption{
15$\mu m$ local differential luminosity function.
}
\end{figure}

The algorithm and the formulation used in this work are the same
as presented in Section 3 of Xu et al. (1998)\footnote{
There was an error in Eq(14) of Xu et al. (1998), which should have been
$Covar(F_{i-1},F_i) = F_i\times Var(F_{i-1})/F_{i-1}$.
}. 
The 60$\mu m$ LLF derived by Saunders et al. (1991)
using the so-called 'non-parametric maximum-likelihood' method is taken,
for which the effects of spatial galaxy density fluctuations,
in particular the local over-density due to the local super cluster, 
are minimized (Saunders et al. 1991). In Table 2
the derived 15$\mu m$ LLF is listed, with $L_{15\mu m}$
being defined by $\nu L_{\nu}$ at 15$\mu m$ and bin width 
$\delta \log (L_{15\mu m}) =0.4$. In Fig.3 this new 15$\mu m$ LLF
is compared to the 15$\mu m$ LLF of Xu et al. (1998) which is derived
from the {\it predicted} $15\mu m$ luminosities of a 25$\mu m$ selected
sample of IRAS galaxies. Excellent agreement, in particular near
the knee of the LLF ($\sim 10^{9.5}$ L$_\sun$), is found between these two 
LLFs, which are derived from completely different data sets using very
different approachs. This verifies that both LLFs are reliable within the 
limits of their uncertainties. On the other hand, the two LLFs are 
complementary to each other. While the new LLF is derived
from {\it real} $15\mu m$ ISOCAM data obtained by Aussel et al.
(2000), the size of this data set (64 galaxies) is much smaller 
than the IRAS sample (1406 galaxies) used by Xu et al. (1998).
Consequently, the new LLF does not extend beyond $L_*$ as far as 
the LLF of Xu et al. (1998),
namely being truncated at 10$^{11.3}$ L$_\sun$ and with the point 
at $10^{10.9}$ missing since there is no galaxy in that bin. 
It should be noted that both the 60$\mu m$ LLF of Saunders et al. 
(1991) on which this work is based, and the 25$\mu m$ LLF of Shupe et al.
(1998) on which the 15$\mu m$ LLF of Xu et al. (1998) is based,
are derived using the maximum-likelihood method which minimizes
the effect of density fluctuations. Also both the normalizations of
the 60$\mu m$ LLF of Saunders et al. (1991) and the 25$\mu m$ LLF 
of Shupe et al. (1998) are carefully determined, and are transferred
by the bivariate analyses to the 15$\mu m$ LLFs of Xu et al. (1998)
and of this work, respectively. The very good agreement between
the points of the two 15$\mu m$ LLFs near the knee (Fig.3) demonstrates
that both normalizations are indeed reliable, eliminating a
large uncertainty in the prediction of the local 15$\mu m$ counts
(Elbaz et al. 1999). In what follows we will use the LLF of Xu et al. (1998) 
when a 15$\mu m$ LLF is needed.

\section{Galaxy Evolution Indicated by ISOCAM 15$\mu m$ Counts} 
In Fig.4 we reproduce the results of ISOCAM 15$\mu m$ surveys presented
in Elbaz et al. (1999). In addition, 
counts derived based on the {\it predicted} 
15$\mu m$ flux densities of sources in the IRAS
25$\mu m$ selected sample of Xu et al. (1998)
are plotted at flux density levels $>0.2$ Jy.
Counts fainter than 0.2 Jy are not plotted since they drop dramatically,
indicating the increasingly severe incompleteness, due to the fact that
the sample is 25$\mu m$ selected rather than 15$\mu m$ selected. Note
that at the bright end ($\geq 0.5$Jy) the normalized counts of these sources
are significantly higher than the counts in the fainter flux density
bins. This excess of counts is very likely due to the overdensity
associated with the local supercluster 
(Lonsdale et al. 1990; Saunders et al. 1991).
\begin{figure}
\plotone{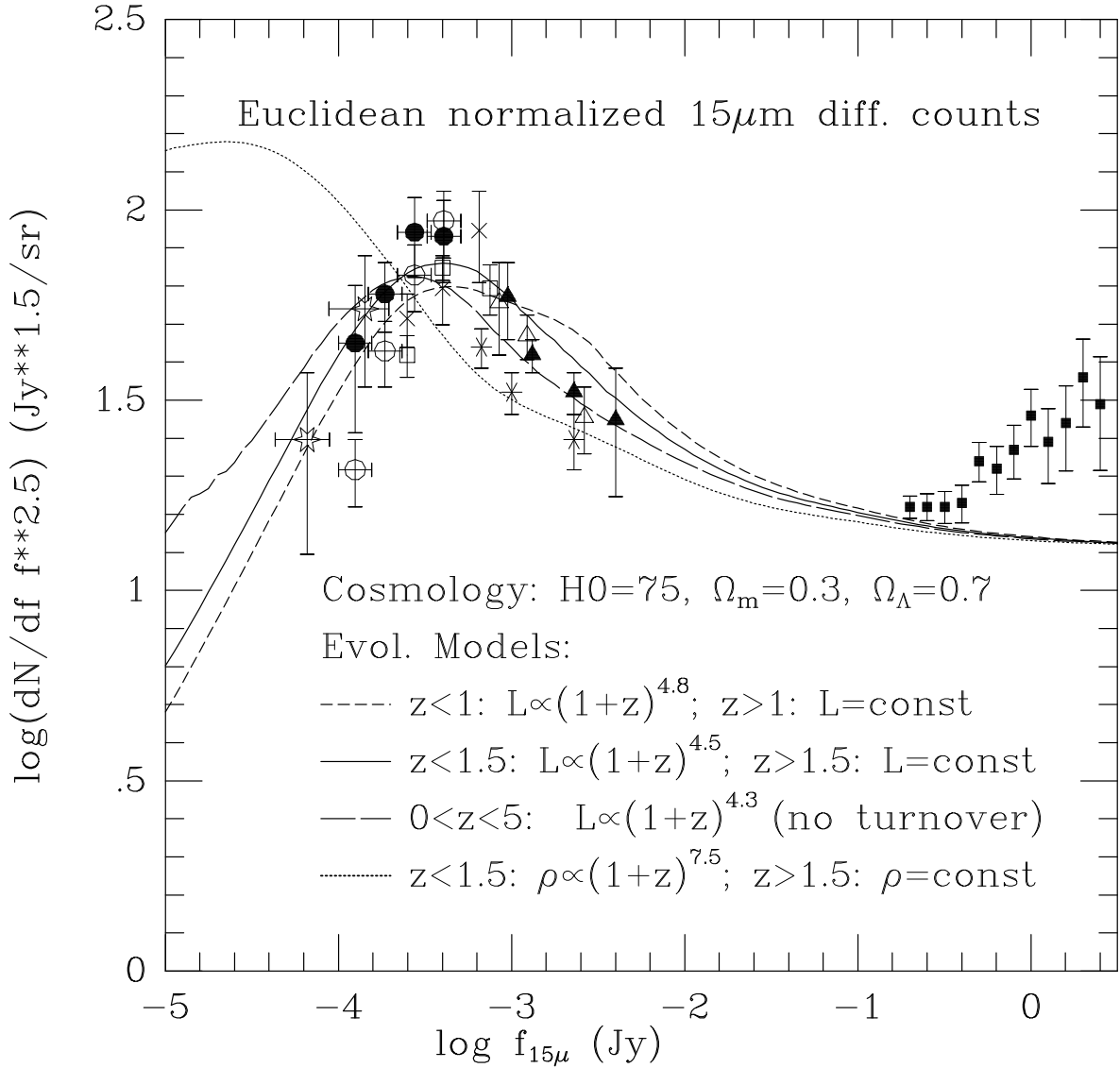}
\vskip-4truecm
\caption{
Euclidean normalized 15$\mu m$ differential
counts: model predictions compared to the observations. The 
observational data are taken from Elbaz et al. (1999). Data points (same
as in Elbaz et al. 1999): A2390 (stars); ISOHDF-North (open circles),
ISOHDF-South (filled circles), Marano FIRBACK Ultra-Deep (open squares),
Marano Ultra-Deep (exes), Marano FIRBACK Deep (stars),
Lockman Deep (open triangles), Lockman Shallow (filled triangles).
At the bright end ($>0.2 Jy$) plotted (filled
squares) are the counts derived
based on the {\it predicted} 15$\mu m$ flux densities of the
sources in the sample of Xu et al. (1998).
}
\end{figure}

As pointed out by Elbaz et al. (1999), the Euclidean normalized 
ISOCAM 15$\mu m$ counts have a narrow and prominent peak at about 0.4 mJy.
There have been suggestions that this peak indicates a new population of
infrared sources emerging after redshift $z\sim 0.4$ (Elbaz et al. 1998a). 
On the other hand
Xu et al. (1998) argued that such a feature can be caused by the broad-band
emission features often associated to the polycyclic aromatic hydrocarbon 
molecules (PAH features, see Puget and Leger 1989). 
However the model predictions published by Xu et al. (1998),
specified by two galaxy evolution models including one
 pure luminosity evolution model
with evolution rate as $L=L_0\times (1+z)^3$ and one pure density
evolution with $\rho=\rho_0\times (1+z)^4$, 
calculated using the number count model which takes 
into account the effects of
these emission features, underestimated the counts when compared with the
ISOCAM data (Elbaz et al. 1998a). 
This suggests that the evolution endured by the
ISOCAM sources is stronger than that assumed by Xu et al. (1998) which is based
on previous studies of IRAS counts (Lonsdale et al. 1990; Saunders et al. 1991;
Pearson and Rowan-Robinson 1996).

Here we present new model predictions using the same number count 
model of Xu et al. (1998), but with stronger evolution rates, and compare 
them with the ISOCAM 15$\mu m$ data. We have assumed that galaxy formation
starts at z=5 (the counts are not sensitive to this parameter).
The cosmology adopted in the models plotted in Fig.4 
is specified by $\Omega_m = 0.3$ 
and $\Omega_\Lambda = 0.7$, which is suggested by recent observations of
type I supernovae in distant galaxies (Garnavich et al. 1998). 
Models with $\Omega_m = 1$ 
$\Omega_\Lambda = 0$ and with $\Omega_m = 0.3$ 
$\Omega_\Lambda = 0$ were also
calculated, but not plotted here. The results from the
$\Omega_m = 0.3$, $\Omega_\Lambda = 0$ cosmology are very close to the results
presented in Fig.4, while the results from the  $\Omega_m = 1$, 
$\Omega_\Lambda = 0$ model fit the data slightly less well. 

The solid line represents
the pure luminosity evolution model with L$\propto (1+z)^{4.5}$ and with
a turnover at $z=1.5$ beyond which the evolution turns flat
(i.e. L=constant for $z\ge 1.5$). The short-dashed line gives
the counts predicted by another pure luminosity evolution model 
with L$\propto (1+z)^{4.8}$ and with
a turnover at smaller redshift: z=1 (i.e. L=constant for $z\ge 1$). 
Finally the long-dashed line is the prediction for counts by a pure luminosity
evolution model without any turnover, and with an evolution rate of
L$\propto (1+z)^{4.3}$.  Among the three luminosity evolution
models, the one with turnover at z=1.5 (the solid curve) gives the best fit, 
closely reproducing the overall shape and the level of the observed counts.
The model with a turnover redshift z=1 predicts a peak which is
too flat compared to the
data. This is because, in the framework of the
Xu et al. (1998) model, the broadband 
MIR features in the wavelength range 6 -- 8.5$\mu m$, which are redshifted
into the ISOCAM LW3 bandpass when z=1$\pm 0.2$, are largely responsible for the
narrow peak of the 
counts in Fig.4. When the turnover occurs at z=1, a significant number
of sources in the redshift range of z=1 -- 1.2 are dropped compared to the models
without turnover or with the turnover at 1.5, resulting in a less prominent peak.
The model without turnover (the long-dashed line) predicts a peak at f$_{15\mu m}
\sim 0.2$ mJy instead of 0.4 mJy as shown by the data.

The dotted line shows the prediction by a density evolution model 
with comoving density $\rho \propto (1+z)^{7.5}$ until z=1.5,
turning flat afterwards (i.e. $\rho$=constant when $z\geq 1.5$).
This model gives a reasonable fit to data points brighter than f$_{15\mu m}
\sim 0.2$ mJy. However in the fainter flux levels,
instead of turning down, the model prediction 
keeps rising in the plot until f$_{15\mu m}$ reaches as low as $\sim 0.04$ mJy.
This is very different from the trend shown by the data.
The reason for the difference between the density evolution model and
the luminosity evolution models
is that for a given redshift, say z=1 which allows the 6 -- 8.5$\mu m$
emission features to be included in the LW3 bandpass, galaxies are much 
fainter in pure density evolution models compared to those in pure
luminosity evolution models, therefore the bump caused by the K-correction
due to the MIR emission features occurs in much fainter flux density 
levels (see Xu et al. 1998 for a more detailed discussion).
\begin{figure}
\plotone{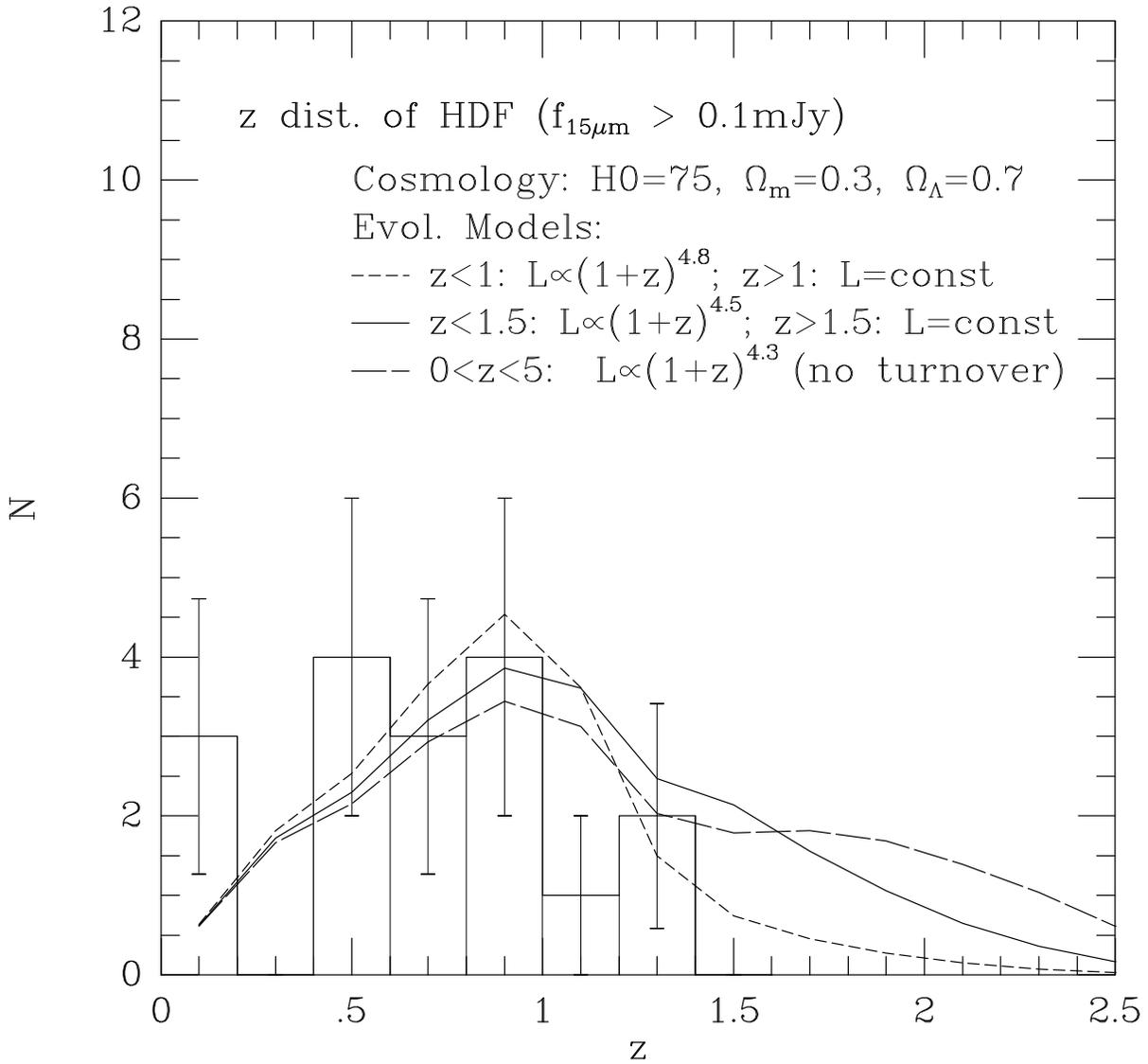}
\vskip-4truecm
\caption{
Redshift distribution of ISOCAM 15$\mu m$ sources in the 
field of ISOHDF-North (Aussel et al. 1999):
model predictions compared to the observational data.
}
\end{figure}

Strong constraints on galaxy evolution can be obtained when the redshift 
information of a flux-limited sample is available. Aussel et al. (1999) found
in the literature 29 redshifts for 49 sources in the main source list of 
the ISOCAM HDF (North) survey. In Fig.5 the histogram of the 
redshifts of 17 of these sources with $f_{15\mu m}\geq 0.1$ mJy
is compared to the model predictions
by the three models presented in Fig.4. 
A sky coverage of 16 acrmin$^2$ (Aussel et al 1999) and a correction factor
of 2 for the incompleteness are 
assumed in these calculations. All three luminosity evolution 
models give reasonably good
fits to the data in the bins with $z<1$, and over-predict the counts
in bins of $z>1$. In particular, 
both the model with turnover at z=1.5 and the model without turnover predict
some sources (6 by the former and 10 by the latter) with z$>1.4$ while none are
found in the data. For the model with turnover at z=1.5, the best
fitting model in Fig.4, the missing
of sources at $z>1.4$ could be due to small number statistics
or to the incompleteness of the data at high redshifts. 
This highlights the demand for larger and more complete redshift samples 
of ISO
sources. In fact, when redshifts for a large (a few hundred) and complete
flux limited sample are available, luminosity functions of ISO sources can be
calculated for different redshift epochs which will give the most 
direct information about the evolution of these sources.

The reasonably good fits to both the source counts (Fig.4) and the
redshift distribution (Fig.5) by the three luminosity evolution models
(in particular the model of L$\propto (1+z)^{4.5}$ with
a turnover at $z=1.5$) demonstrate that indeed the narrow peak
of the ISOCAM 15$\mu m$ counts can be well explained by the effect of
broadband MIR emission features which is the essential element of the
model of Xu et al. (1998), and there is no need to invoke a 'new
population'. At the same time, the evolution rate implied by the model
fit is much stronger than those given by
previous studies on IRAS sources (L$\propto (1+z)^3$, Pearson and
Rowan-Robinson 1996; Lonsdale et al. 1990), but is consistent with
what is found in the UV and optical deep surveys (L$\propto
(1+z)^{3.95\pm 0.75}$, Lilly et al. 1996). Analyzing 
multiband data from IRAS, ISO, SCUBA and COBE, Blain et al. (1999a)
also obtained a relatively high evolution rate ((L$\propto
(1+z)^{3.8\pm 0.2}$). Given the lack of
dependence of the $L_{15\mu m}/L_{60\mu m}$ ratio on the luminosity, 
one expects that the evolution rates of the $L_{15\mu m}$ and
of the $L_{60\mu m}$ should be similar. In Fig.6 we compare the 
counts at 60$\mu m$ predicted by a luminosity evolution model 
assuming $L_{60\mu m}\propto (1+z)^{4.5}$ which turns flat
($L_{60\mu m}=$constant) at $z=1.5$, with IRAS data.
The large filled circles are the counts from this work (new NEPR sample),
which are about 30\% (i.e. $\sim 0.12$ dex) higher than those of HH87
(crosses).
This discrepancy can be fully explained by the fact that the 60$\mu m$
fluxes obtained by the new SCANPI/SUPERSCANPI processing are
about 20\% higher than those of HH87 (Fig.1a), given that 
the Euclidean normalized differential 
counts scale with the flux to the 1.5 power. 
At the same time, the model 
predictions (solid line) indeed reproduce the trend and the level of
those more recent IRAS counts (Bertin et al. 1997; Gregorich
et al. 1995; Saunders et al. 1991; Lonsdale et al. 1990; Rowan-Robinson
et al. 1990) quite well.
\begin{figure}
\plotone{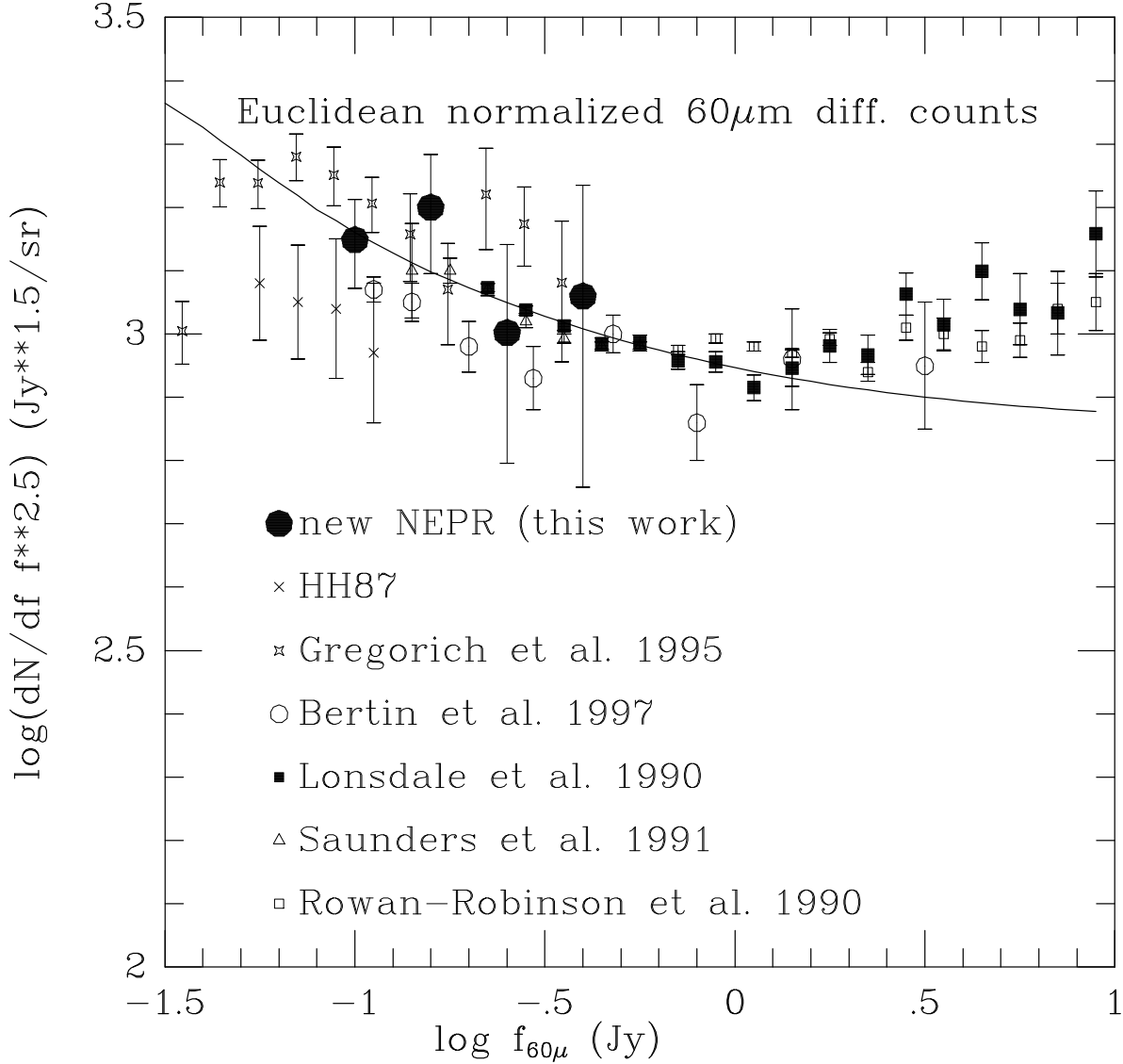}
\vskip-4truecm
\caption{
Euclidean normalized differential
counts of the IRAS 60$\mu m$ band: 
model predictions compared to the observations. 
Data points: new NEPR (this work) (large filled circles); 
HH87 (exes); Gregorich et al. 1995 (open stars);
Bertin et al. 1997 (open circles); Lonsdale et al. 1990 (filled squares);
Saunders et al. 1991 (filled squares), Rowan-Robinson et al. 1990 (open
squares). Model:  $L_{60\mu m}\propto (1+z)^{4.5}$ when $z\leq 1.5$,
$L_{60\mu m}=$constant when $z>1.5$ (solid line).
}
\end{figure}

Pure density evolution, which gives a poor fit to the number counts, 
can be ruled out with high confidence. Given that any
density evolution will push the peak in the number counts to fainter
flux levels than shown by the data, it is quite certain that, as far
as the choice between density and luminosity evolution is concerned, 
luminosity evolution is the dominant cause of the high
number counts. This conclusion 
is in agreement with that of Blain et al. (1999a) which is obtained from a
completely different argument, namely that a pure density evolution 
model which can fit the IR/submm counts will
produce too much IR/submm background.

Our results show that the narrow peak of the ISOCAM 15$\mu m$ counts 
at about 0.4 mJy may not be used as an evidence for a 'new population'
of faint MIR sources. On the other hand, a luminosity evolution in the
luminosity function of infrared galaxies, as suggested by our
best fitting model, does not necessarily mean
that it is the same galaxies that we are seeing in the local Universe
that are shining tens or even hundreds times brighter in the early
epochs of the Universe. Indeed, the
preliminary results of optical identifications of
ISOCAM LW3 sources indicate that beyond $z\sim 0.7$ most of them are 
interacting galaxies (Aussel et al. 1999; Elbaz et al. 1999), while the local
MIR selected extragalactic sources are mostly single late type
galaxies similar to the Milky Way (Rush et al. 1993). 
Given the high incompleteness of the redshift data, at this stage 
the major constraint on the evolution of ISOCAM sources
is from the counts, which is mostly determined by how the luminosity function
evolves around L$_*$ (the luminosity distribution of sources in a given
f$_{15\mu m}$ bin peaks strongly around L$_*$). Therefore,
without any extrapolation,  what we know now is that whatever the
population of IR sources is at redshift $z\sim 1$, its comoving
density is about the same as the IR galaxies in the local Universe,
while the characteristic 15$\mu m$ luminosity of the faint sub-mJy
sources is about 20 times the $L_*$ of local IR galaxies, namely at
the $L_*\sim 10^{11}$ L$_\sun$ level. If these sources are indeed similar
to the local gas-rich spiral-spiral galaxy pair systems, which
dominate the bright end ($L_{fir} > 2\times 10^{11}$L$_\sun$) of IRAS
luminosity function (Xu and Sulentic 1991), the implied density
enhancement of these sources at $z\sim 1$
compared to their density in the local Universe is more than an order
of magnitude. Although the population of these interacting galaxies is not
really 'new' (i.e. they are already important contributors of
MIR counts in the local Universe), it is quite possible that this
population may evolve much faster than normal late-type
galaxies, and even than AGNs.
For the sake of simplicity, we have treated all IR
galaxies as a single population in our model and have not considered
any 'differential evolution' (i.e. different evolution rates for
galaxies with different luminosities).  When more constraints on the
nature of 15$\mu m$ sources are available from future follow-up
observations, a model treating the evolution of different galaxy
populations differently (e.g. separating interacting galaxies from single
galaxies and AGNs), such as the model by Franceschini et
al. (1994), will be more appropriate.

Our results also suggest that the MIR emission
features are present in the SEDs of galaxies with redshifts
up to $z\sim 1$. Whether 
this is still true for galaxies with even larger redshifts will
be found out by the future SIRTF mission (Cruikshank and Werner 1997).
If so, these features will facilitate a powerful
new method of obtaining redshifts in infrared for the optically faint, heavily 
extinguished galaxies.

As for the question of whether galaxy
evolution has a turnover at z=1--2, our results are not
conclusive, though a positive answer is favoured by the model fits (Fig.4).
Given the significant effect of the MIR emission features, which happens to
affect the 15$\mu m$ counts at redshift just below z=1.5, the turnover favoured
by our model may well be a false signal. Deep surveys at longer wavelengths
(e.g. at 25$\mu m$ and 70$\mu m$ using the SIRTF/MIPS detector arrays)
where the MIR features will be redshifted into the bandpass at larger z,
are certainly desirable for the determination 
of the evolution of IR galaxies beyond z=1.5 (Xu et al. 1998).
\begin{figure}
\plotone{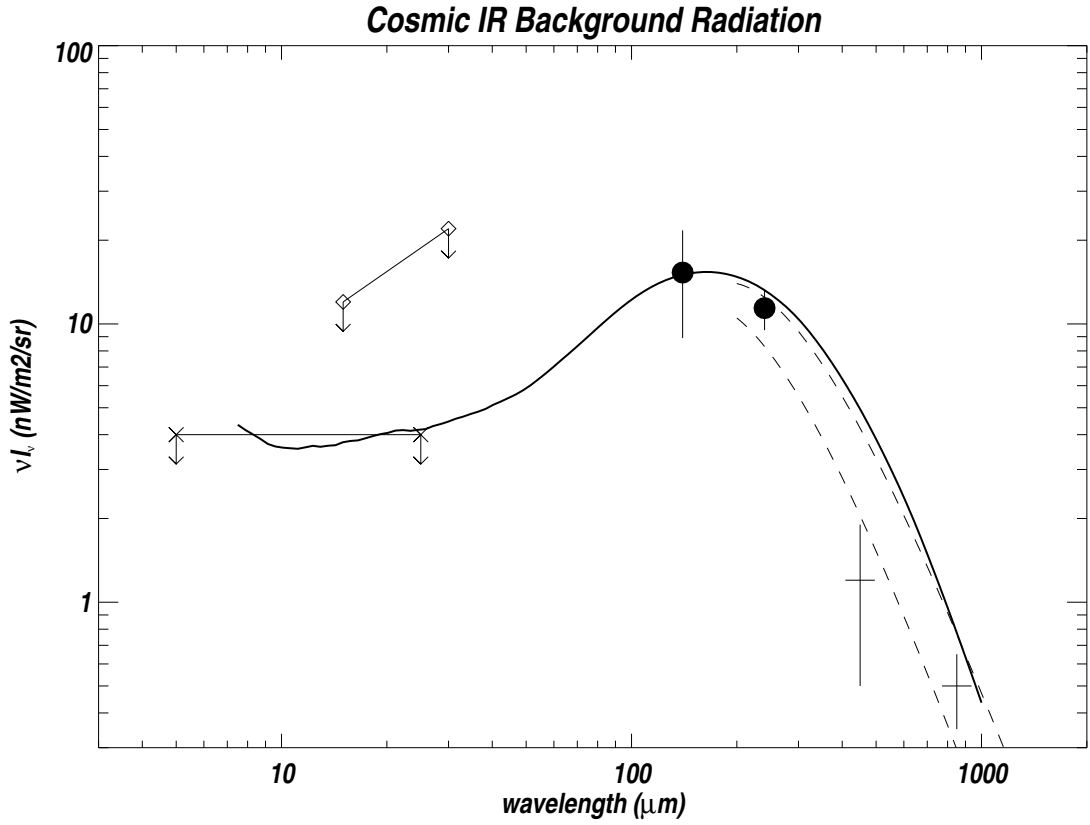}
\vskip-6truecm
\caption{
Cosmic IR/submm background:  
model predictions compared to the observational data.
Symbols: Model prediction ($z \leq 1.5$: L$\propto (1+z)^{4.5}$; $1.5 < z \leq
5$: L=constant) --- solid curve; COBE/DIRBE
results (Lagache et al. 1998) --- large filled circles with error bars;
SCUBA results (Blain et al. 1999) --- large crosses;
the range of COBE/FIRAS results (Fixen et al. 1998) ---
two dashed curves; upperlimits 
from TeV gamma-ray radiation of Mrk403 and Mrk501
 (Dwek \& Slavin 1994; Stanev \& Franceschini 1998) --- diamonds and 
exes with upper-limits.}
\end{figure}

How much do these sources contribute to 
the Cosmic IR/sub-mm Background
radiation? To answer this question, a Monte Carlo
simulation based on the source count
model of Xu et al. (1998) was carried out, assuming the galaxy evolution model
that gives the best fit to the data in Fig.4 
(L$\propto (1+z)^{4.5}$ with
a turnover at $z=1.5$). Sources with certain $L_{15\mu m}$ and z are
generated according to predictions of the number count model, and IR SEDs
taken from the SED sample of 1406 galaxies in Xu et al. (1998) are assigned
to these sources in accordance with their rest frame $L_{15\mu m}$. 
It should be noted that the SEDs modeled by Xu et al. (1998) 
stop at 120$\mu m$.
In order to estimate the contribution from MIR galaxies to the IR background
radiation at longer wavelengths, we have assumed that the IR emission at
$\lambda > 120\mu m$ of all sources
generated in the simulation has the same spectrum specified 
by a modified blackbody
with $T=40$K and the emissivity index of $\beta=1.5$ (Blain et al. 1999). 
In reality, this submm SED may only apply to the luminous IR
starburst galaxies (LIRGs) while the SEDs of 
less active galaxies are likely to be much colder (Eales et al. 1999).
However, since the largest contribution to the submm background is
from LIRGs (Blain et al. 1999), we neglect this complication here.
 
For any given wavelength, all the fluxes from these simulated sources
are summed up, resulting in the predicted contribution of the
population of $15\mu m$ sources at the wavelength in question. Again
we assume that galaxy formation starts at z=5. In Fig.7, this
prediction (the solid curve) is plotted against several
measurements/upper-limits of the IR/submm background. The upper-limits
are all from the studies of TeV gamma-ray sources (Dwek \& Slavin
1994; Stanev \& Franceschini 1998).  The filled circles with error
bars are COBE/DIRBE results taken from Lagache et al. (1998), and the
large crosses are from SCUBA results (Blain et al. 1999).  The two
dashed curves outline the range of the submm IR background detected by
COBE/FIRAS (Fixen et al. 1998). According to our results, the
contribution of the MIR galaxies to the background between 10--30$\mu
m$ is at the 4 nW/m$^2$/sr level, compared to the results reported by
Elbaz et al. (1999) that the 15$\mu m$ background due to sources brighter
than 50$\mu$Jy is 3.3 nW/m$^2$/sr. Note that this already meets the
upper-limits obtained by Stanev and Franceschini (1998) from the
analysis of TeV emission of Mrk501. At longer wavelengths, the predicted
contribution to the background emission agrees very well with the DIRBE
points, and lies slightly above the upper-boundary of the measured submm
background. Compared to the results of previous calculations on the
cosmic IR background using `backward evolution' models (Hacking and
Soifer 1991; Beichman and Helou 1991; Malkan and Stecker 1998), the
category this work belongs to, our result is about a factor of 2
higher because the evolution rate hinted at by ISOCAM 15$\mu m$ surveys
is significantly stronger than those used in the previous works. Our
result is in agreement with Elbaz et al (1999), who found that the ISOCAM
15$\mu m$ sources may be able to account for the majority of the IR/submm
background.  Taken at face value, the result in Fig.7 indicates
that nearly all of the sources contributing significantly to the IR/submm
background are already present in the population of 15$\mu m$ sources
detected by ISO, and very little room is left for any missing 'new population' 
which can be significant energy sources of the IR/submm sky.


\vskip2cm 
The author is indebted to Gianfranco De Zotti, who not only
suggested this project, but also provided stimulating 
comments on an earlier version of this paper. He also thanks
Herve Aussel and David Elbaz for informative and constructive discussions.
Diane Engler, Iffat Khan, Joe Mazzarella and Steve Lord
are thanked for helping to obtain and interpret the SCANPI/SUPERSCANPI 
results. David Gregorich and Perry Hacking are thanked
for providing information on IRAS pointed observations in NEPR.
Constructive comments on the manuscript 
from Carol Lonsdale, David Shupe and Andrew Blain, 
and from an anonymous referee 
are acknowledged. Part of the work is supported by 
NASA grant for ISO Data Analysis. 
This research has made use of the NASA/IPAC Extragalactic Database (NED) 
which is operated by the Jet Propulsion Laboratory, California Institute of 
Technology, under contract with the National Aeronautics and
Space Administration. The author 
was supported by the Jet Propulsion Laboratory,
California Institute of Technology, under contract with NASA.


\clearpage

\begin{deluxetable}{cccccc}
\tablewidth{0pt}
\tablecaption{New f$_{60\mu m}$ for the NEPR sample}
\tablehead{
\colhead{I.D.\tablenotemark{1}} 
& \colhead{f$_{60\mu m}$\tablenotemark{2}} 
& \colhead{error\tablenotemark{3}}
& \colhead{f$^{HH}_{60\mu m}$\tablenotemark{4}} 
& \colhead{offset}\tablenotemark{5} &
\colhead{redshift\tablenotemark{6}}  \nl
\cline{1-6} \nl
\colhead{(HH87)}
& \colhead{(mJy)}
& \colhead{(mJy)}
& \colhead{(mJy)}
& \colhead{($''$)}
& \colhead{ }
}
\startdata 
3-01     &    &  27    &     83  &   81.5  &   0.116   \nl 
3-02     &  80 &  16   &     85  &  104.8  &    	   \nl 
3-03     &  140 &  17  &    100  &    6.6  &   0.089   \nl 
3-04     &  250 &  20  &    190  &    7.1  &   0.121   \nl 
3-05     &  85 &  18   &    100  &   11.3  &   0.0408  \nl 
3-07     &  110 &  14  &     78  &    7.4  &   0.0417  \nl 
3-08     &  220 &  13  &    170  &    1.6  &   0.052   \nl 
3-09     &   &  25     &     73  &   17.6  &   0.0255  \nl 
3-10     &  330 &  21  &    260  &    8.3  &   	   \nl 
3-11     &  170 &  22  &    110  &    7.4  &   0.0250  \nl 
&&&&& \nl
3-12     &  150 &  17  &    130  &   23.4  &   0.0766  \nl 
3-13     &  64 &  25   &     66  &   29.2  &   0.201   \nl 
3-14     &  130 &  12  &     99  &   13.5  &   0.0421  \nl 
3-15     &  130 &  23  &    110  &   14.3  &   0.0780  \nl 
3-16     &  90 &  14   &     92  &   20.9  &   0.117   \nl 
3-17a    &   &  16     &         &   42.1  &    \nl 
3-17b    &  60 &  14   &     63  &   12.8  &   0.0704  \nl 
3-18     &  140 &  25  &     82  &   56.3  &   0.229   \nl 
3-19abc  &  130 &  17  &     74  &   30.6  &   0.0872  \nl
3-20ab   &  103 &  17  &     62  &   52.3  &   0.0735  \nl 
&&&&& \nl
3-21     &  60 &  10   &     64  &   17.8  &   0.0522  \nl 
3-23     &  90 &  22   &     87  &   15.0  &   0.0878  \nl 
3-24     &  90 &  13   &     74  &    9.1  &  	   \nl 
3-25     &  100 &  20  &     77  &   20.7  &   0.0853  \nl 
3-26abcd &  150 &  10  &    140  &   21.3  &   0.089   \nl 
3-27a    &  110 &  12  &     87  &   26.1  &   0.0873  \nl 
3-27b    &  70 &  14   &         &   55.6  &   0.0873  \nl 
3-28     &  80 &  16   &     63  &   21.4  &   0.0253  \nl 
3-29     &  50 &  14   &     57  &   24.3  & 	   \nl 
3-30     &  70 &  21   &     72  &   41.0  &   0.2540  \nl 
&&&&& \nl
3-31     &  170 &  12  &    190  &    5.8  &   	   \nl 
3-32a    &  90 &  21   &     89  &   43.6  &   0.0375  \nl 
3-32b    &  30 &  13   &         &   95.8  &   0.0375  \nl 
3-33a    &  70 &  12   &     56  &   24.0  &   0.1440  \nl 
3-33b    &   &  12     &         &   72.3  &    \nl 
3-34     &  100 &  14  &     89  &    9.0  &   0.0373  \nl 
3-35     &  70 &  14   &     70  &   31.5  &   0.195   \nl 
3-36a    &   &  71     &     66  &    8.2  &   0.119  \nl 
3-36b    &   &  72     &         &   69.9  &      \nl 
3-37     &  480 &  16  &    510  &    2.0  &   	   \nl 
&&&&& \nl
3-38ab   &  80 &  10   &     54  &   56.2  & 	   \nl 
3-39     &  80 &  26   &     57  &   14.2  & 	   \nl 
3-40ab   &  110 &  26  &     98  &   27.7  &   0.0887  \nl 
3-41     &  100 &  19  &    100  &   15.2  & 	   \nl 
3-42     &  50 &  16   &     60  &   39.1  &   0.1150  \nl 
3-43     &  140 &  16  &    150  &   18.6  &   0.026   \nl 
3-44     &  080 &  17  &     54  &   24.4  &   	   \nl 
3-45     &  160 &  18  &    140  &    6.7  &   0.0789  \nl 
3-46     &  80 &  15   &     71  &    2.9  &   0.0360  \nl 
3-47     &  160 &  22  &    110  &   28.9  &   0.0867  \nl 
&&&&& \nl
3-48     &  160 &  17  &    140  &   23.6  &   0.0259  \nl 
3-49     &  50 &  16   &     54  &   14.6  & 	   \nl 
3-50     &  100 &  26  &     89  &    6.7  &   0.0876  \nl 
3-51a    &  160 &  11  &    130  &    9.0  & 	   \nl 
3-51b    &   &  18     &         &   96.2  & 	   \nl 
3-53     &  400 &  29  &    330  &    3.2  &   	   \nl 
3-54     &  80 &  16   &     74  &   14.6  &   0.0799  \nl 
3-55     &  70 &  22   &     54  &   15.6  &   	   \nl 
3-56     &  110 &  19  &    110  &   10.7  &   0.0881  \nl 
3-57     &  230 &  11  &    220  &   11.3  &   	   \nl 
&&&&& \nl
3-58     &  80 &  13   &     50  &   14.7  & 	   \nl 
3-59     &  70 &  12   &     57  &    9.1  & 	   \nl 
3-61     &  160 &  15  &    150  &    6.0  & 	   \nl 
3-62     &  110 &  20  &     83  &   12.0  &   0.0271  \nl 
3-63a    &  210 &  28  &    210  &    6.3  &   0.1180  \nl 
3-63b    &   &  13     &         &  104.3  &  	   \nl 
3-64     &  560 &  20  &    540  &    3.6  & 	   \nl 
3-65     &  110 &  11  &    120  &   18.7  &   0.173   \nl 
3-66     &  96 &   8   &     89  &   10.7  &   0.0535  \nl 
3-67ab   &  110 &  11  &     97  &   23.5  &   0.0399  \nl 
&&&&& \nl
3-68     &  50 &  21   &     51  &   35.6  & 	   \nl 
3-69     &  50 &  20   &     61  &   15.2  &   0.104   \nl 
3-70     &  140 &  10  &    100  &    8.4  &   0.197   \nl 
3-71     &  80 &  13   &     72  &    3.1  &   0.0517  \nl 
3-72     &  80 &  30   &     57  &   75.4  &   	   \nl 
3-73     &  230 &  13  &    250  &   13.6  & 	   \nl 
3-74     &  150 &   8  &    160  &   14.9  &   0.026   \nl 
3-75     &   &  19     &     65  &   23.6  &   0.0583  \nl 
3-76     &  80 &  16   &     61  &   20.7  &   0.0800  \nl 
3-77     &  100 &  16  &    120  &   57.1  &   0.0871  \nl 
&&&&& \nl
3-78a    &   &  12     &         &   68.1  &  	   \nl 
3-78bc   &  90 &  23   &     69  &   13.1  &   0.0774  \nl 
3-79ab   &  300 &  11  &    230  &   19.6  &   	   \nl 
3-79c    &   &  10     &         &  110.7  &   	   \nl 
3-80ab   &  330 &  20  &    290  &   13.7  & 	   \nl 
3-81ab   &  130 &  14  &    110  &   23.4  &   0.0268  \nl 
3-82     &  220 &  13  &    200  &    6.1  &   0.0551  \nl 
3-83     &  100 &  14  &     89  &   15.7  &   0.107   \nl 
3-84     &  160 &  11  &    160  &    5.8  &   0.086   \nl 
3-85     &  100 &   9  &     69  &   19.3  &   0.151   \nl 
&&&&& \nl
3-86     &  70 &  16   &     73  &   31.2  &   	   \nl 
3-88a    &  100 &  17  &     74  &   29.0  &   0.519   \nl 
3-88b    &   &  17     &         &   86.9  &      \nl 
3-89     &  90 &  16   &     69  &   56.5  &   0.029   \nl 
3-90     &  100 &   9  &     65  &    8.6  &   0.0720  \nl 
3-91     &  80 &  11   &     58  &   15.2  & 	   \nl 
3-92abc  &  90 &  19   &     75  &   56.9  &   0.0880  \nl 
3-93     &  140 &  19  &     99  &   23.0  &   0.0690  \nl 
3-94a    &  130 &  16  &     98  &    3.7  &   0.0502  \nl 
3-94b    &   &  20     &         &   89.3  &  	   \nl 
&&&&& \nl
3-96     &  190 &  18  &    150  &    5.8  &   0.0321  \nl 
2-16     &  40 &  13   &     38  &    9.5  & 	   \nl 
3-06     &    &        &     62  &         & 0.0754  \nl 
3-22     &   &         &     88  &         &    \nl 
3-87     &   &         &     55  &         &    \nl 
3-95     &   &         &     78  &         &  0.187   \nl 
&&&&& \nl
\enddata
\tablenotetext{1}{These are the same sources
as listed in Table 3 of Aussel et al. (2000). 
Sources confused with each other in the 60$\mu m$ band are grouped
together as single entries.}
\tablenotetext{2}{60$\mu m$ flux density obtained in this work.
For undetected sources the entry is blank.}
\tablenotetext{3}{The one-$\sigma$ error of $f_{60\mu m}$.}
\tablenotetext{4}{60$\mu m$ flux density taken from HH87.
When more than one source corresponds to a single HH87 source,
the flux density is assigned to the one with the least offset.}
\tablenotetext{5}{Offset of the 15$\mu m$ source from the HH87 source,
taken from Aussel et al. (2000).
When more than one 15$\mu m$ source corresponds to a single source
here (confused sources), the smallest offset is taken.}
\tablenotetext{6}{Redshift taken from Ashby et al. (1996).}
\end{deluxetable}

\begin{deluxetable}{ccc}
\tablewidth{0pt}
\tablecaption{15$\mu m$ Local Luminosity Function (LLF) from the NEPR sample
\label{tab:lf15}
}
\tablehead{\colhead{$\log(\nu L_{\nu} (15\mu m)/L_\sun )$} 
& \colhead{$\rm \log(\phi /(Mpc^{-3}mag^{-1}))$} & \colhead{1 $\sigma$ error}
}
\startdata
     8.1 &  -3.70  &   -3.71 \nl
     8.5 &  -2.42  &   -2.78 \nl
     8.9 &  -2.68  &   -3.40 \nl
     9.3 &  -3.09  &   -3.56 \nl
     9.7 &  -3.47  &   -4.21 \nl
     10.1 &  -4.11  &   -4.43 \nl
     10.5 &  -5.08  &   -5.64 \nl
     11.3 &  -7.49  &   -7.49 \nl
\enddata      
\end{deluxetable}

\end{document}